\documentclass[12pt]{article}
\newcommand{\etal}{{\it et al.}}

\newcommand{\average}[1]{\mbox{$\displaystyle\left\langle#1\right\rangle$}}

\newcommand{\eq}[1]{(\ref{#1})}
\newcommand{\PLB}[3]{Phys.~Lett.~{\bf B~#1} (#2)~#3}
\newcommand{\PLBc}[3]{{\bf B~#1} (#2)~#3}
\newcommand{\ZPC}[3]{Z.~Phys.~{\bf C~#1} (#2)~#3}
\newcommand{\PRL}[3]{Phys.~Rev.~Lett.~#1 (#2)~#3}
\newcommand{\PRD}[3]{Phys.~Rev.~{\bf D~#1} (#2)~#3}
\newcommand{\NP}[3]{Nucl.~Phys.~{\bf #1} (#2)~#3}

\newcommand{\ii}{\'\i}

\newcommand{\jpsi}{$J/\psi$\ }

\usepackage{citesort}
\usepackage{amstex}
\usepackage{graphicx}

\begin{document}
\begin{titlepage}

\title{\vskip 3truecm Absorption and J/$\psi$ Suppression in \\ Heavy Ion Collisions}
\author{J. Dias de Deus \\ Dept. of Physics and CENTRA, IST, Lisbon, Portugal \\
\and 
        J. Seixas\thanks{On leave from Dept. F\ii sica, Instituto Superior T\'ecnico, Av. Rovisco Pais, 1096 Lisboa Codex,
        Portugal}\\ Theory Division CERN \\ CH 1211 Geneva 23, Switzerland}
\begin{abstract}
We discuss the \jpsi suppression in the framework of multiple collision models. From the analysis of the Pb--Pb NA50 data we
conclude that the strength of the absorption has increased, but we find no clear evidence for the formation of the quark--gluon
plasma.
\end{abstract}
\maketitle
\end{titlepage}
\newpage
The suppression of the \jpsi production in nucleus--nucleus collisions was proposed, more than ten years ago, by Matsui and
Satz \cite{Matsui Satz} as a signal for the formation of the quark--gluon plasma (QGP). The suggested mechanism is simple: when
colour is liberated Debye screening prevents the creation of the $c\bar{c}\to J/\psi$ bound state.

Suppression of the \jpsi relative to Drell--Yan (DY) production was indeed observed by the NA38 collaboration in O--U and S--U
collisions \cite{NA38}. However, it was soon realized that conventional nuclear absorption of the \jpsi should play, qualitatively
at least, a similar role \cite{Capella}. The problem of disentangling absorption effects from quark--gluon plasma formation became a
central issue in the field of heavy-ion collisions.

In recent years, a great deal of theoretical effort was developed in trying to clarify the origin of a large \jpsi absorption that
would be able
to justify the observed suppression. In fact, a cross--section of the order of 7 mb, much larger than the observed $\psi N$
cross--section, is needed\cite{Carlos}. In this context, the role of interacting pre--resonant $c\bar{c}$ states in matter is to help achieve such
large cross--section \cite{Bodwin}.

In the last year the NA50 collaboration has reported an anomalous \jpsi suppression in Pb--Pb collisions \cite{NA50}. This anomaly
appears as a deviation from the conventional plot of the ratio \jpsi over DY against the average path length $L$ for large values of
the transverse energy $E_T$. This average path length is used in the parametrization $R(L)$ of the \jpsi over DY suppression 
\begin{equation}
R(L)=\exp(-\rho\sigma L) \label{definition of L}\qquad ,\label{L definition}
\end{equation}
$\rho$ being the nuclear density and $\sigma$ the absorption cross--section.

Several explanations exist for the anomalous suppression. Some are based on the quark--gluon plasma formation \cite{Blaizot}. Others
on the additional destruction of the \jpsi in the interaction with comovers formed in the collision \cite{Gavin}. In the latter
case, the effect amounts to an increase of $\rho$ with $E_T$ and effectively to a decrease of the \jpsi over DY ratio faster than
exponential. A discussion of these explanations is contained in  \cite{Carlos}. 

More recently, the NA50 collaboration, from the analysis of the 1996 data, reported in addition to the anomalous suppression, a
discontinuity or abrupt fall in the ratio \jpsi over DY around $E_T\simeq 50$ GeV \cite{Ramello}.

Next we would like to look at the NA50 data from the point of view of a very simple model \cite{DD1,DD2}. In principle, the model
should be able, to some extent, to discriminate quark--gluon plasma formation from absorption effects. The essential features of the
model are the following:
\begin{enumerate}
\item Nuclear collisions are a superposition of elementary (nucleonic, partonic) collisions. This is the framework of the Glauber
model, Dual Parton Model (DPM) and many other models.
\item Fluctuations in multiplicity $n$ and in transverse energy $E_T$ are mainly determined by the fluctuations in the number $\nu$
of elementary collisions. This is required when comparing distributions in nucleus--nucleus collisions with distributions in
nucleon--nucleon collisions.
\item In the production of a rare, unabsorbed event ${\cal C}$, as DY or $W/Z$, the $E_T$ distribution $P_{\cal C}(E_T)$, associated
to the rare event ensemble and the minimum--bias $E_T$ distribution $P(E_T)$ are universally related by
\begin{equation}
P_{\cal C}(E_T)=\frac{E_T}{\average{E_T}}P(E_T)\qquad ,
\end{equation}
or, if normalization cannot be established, by
\begin{equation}
N_{\cal C}(E_T)\sim E_T N(E_T)\qquad ,\label{relation}
\end{equation}
$N_{\cal C}$ and $N$ denoting the number of rare and minimum--bias events, respectively. This relation is (approximately) true in
detailed calculations with multiple--collision models.
\item As absorption means that not all collisions producing event ${\cal C}$ are effective, the effect can be included by making the
change
\begin{equation}
\nu\to\nu^{\epsilon}\quad ,\quad 0\le\epsilon\le 1\qquad ,
\end{equation}
thus effectively decreasing $\nu$. As a consequence, instead of (\ref{relation}), the relation reads
\begin{equation}
N_{\cal C}(E_T)\sim E_T^{\epsilon} N(E_T)\qquad .\label{absorption}
\end{equation}
For the ratio of the absorbed \jpsi over the unabsorbed DY, from (\ref{absorption}) and (\ref{relation}), we arrive at 
\begin{equation}
R(E_T)=N_{J/\psi}(E_T)/N_{DY}(E_T)\sim 1/E_T^{\gamma}\quad ,\quad \gamma=1-\epsilon\qquad .\label{ratio absorption}
\end{equation}
All the absorption models we are aware of, including models with comovers, show for the ratio a behaviour roughly as in 
(\ref{ratio absorption}), in particular with a positive curvature as a function of $E_T$.
\end{enumerate}

The model was successfully tested with the NA38 S--U data. Relations (\ref{absorption}), for DY, and (\ref{relation}), for \jpsi
production with $\epsilon_{J/\psi}\simeq 0.7$, were seen to work fairly well. As 1995 NA50 data on minimum--bias, DY and \jpsi $E_T$
distributions are now available \cite{tese}, we will compare those data with our simple test model.

In Fig. 1 we show the 1995 NA50 experimental points for the $E_T$ distribution associated to the production of Drell--Yan
$\mu^+\mu^--$ pairs. The curve was obtained from the minimum--bias experimental $E_T$ distribution by using relation
(\ref{relation}). The general shape is correctly described. At low $E_T$, since there are experimental efficiency problems, one
does not expect any kind of agreement. To the curve itself there are associated errors (not shown), specially in the low--$E_T$ region,
mainly due to the difference in binning of the minimum--bias and DY experimental distributions.

In Fig. 2 we present the NA50 experimental points for the $E_T$ distribution associated to the production of \jpsi. The solid curve
corresponds to relation (\ref{absorption}) with $\epsilon_{J/\psi}=0.6$. Once more the low--$E_T$ experimental points are not
reliable and one should not expect a good fit.

In Fig. 3a we show the 1995 data for the ratio \jpsi over DY and the parametrization (\ref{ratio absorption}) with
$\gamma_{J/\psi}=1-\epsilon_{J/\psi}=0.4$, while in Fig. 3b we show the same ratio for the 1996 data with the same
parametrization. In both cases the agreement is fairly good.

It is perhaps interesting to try to understand the meaning of the absorption parameter $\epsilon$. The fact that it has decreased
from $\epsilon\simeq 0.7$ in S--U to $\epsilon\simeq 0.6$ in Pb--Pb is an indication of the increase in \jpsi absorption and
destruction in the interacting medium.

If we use the variable $L$, the average path length, and write the \jpsi over DY suppression ratio in the conventional form \eq{L
definition} and compare it with (\ref{absorption}),
\begin{equation}
R(E_T)=E_T^{\epsilon-1}\qquad ,
\end{equation}
we obtain
\begin{equation}
\ln E_T = \frac{\rho}{1-\epsilon}\sigma L\qquad . \label{ETL relation}
\end{equation}
Relation (\ref{ETL relation}) allows, in principle, for the determination of $\epsilon$. Notice that $\epsilon=1$ (no absorption)
requires $\rho=0$ (no medium). It is clear that as absorption increases, $\epsilon$ decreases and, for the same  $E_T$, $L$ increases
as expected\cite{Gonin}.

However (\ref{ETL relation}) cannot be valid for very large $E_T$ since $L$ is finite. The fact that $L$ has an upper bound $L_{max}$ 
implies that looking for \jpsi having an interaction in a range larger than this value is impossible and the ratio $J/\psi$/DY simply
cannot exist beyond it. The $E_T$ variable does not suffer from this problem since, by definition, $E_T$ measures the
existence of a real interation and therefore has to occur inside the interaction region anyway. In this sense $E_T$ is a completely
reliable variable, while $L$ is not.

To understand the sudden drop in the ratio $J/\psi$/DY as a function of $L$ we have to look at what happens as we increase the centrality of the events. In
this case  the density in the interaction region also increases and other degrees of freedom are going to be excited, making it
increasingly opaque. Therefore, as we increase $E_T$ it will be more and more difficult for the \jpsi to escape the interaction
region, leading thus to the observed drop in the ratio $R(L)$. Whether these new degrees of freedom correspond to QGP plasma droplets
or any other object does not come out from the analysis we have just performed. The only thing we can say is that 
deviations from (\ref{L definition})
can be seen as due to the appearence of new degrees of freedom in the physics of nuclear collisions. However, describing them solely as 
signals of QGP is probably pushing a little too far, since the shape of $R(E_T)$ clearly indicates absorption mechanisms at work.

\bigskip\noindent
\textbf{\Large Conclusions}
\bigskip

It is by now clear that the simple test model developed in  \cite{DD1} reproduces rather well the data for the NA38--50 experiments in a
large range of energies and of nuclei. It is true that the model does not predict the form of the minimum--bias distribution but rather
uses this information to describe the DY and \jpsi distributions. Since this model has only one free parameter, namely the
absorption parameter $\epsilon$, and has a very simple, but fundamental set of hypotheses for multicollision of elementary
participants, it serves as a straightforward and powerful way of checking the consistency of the data. It is also interesting to
notice that so far the model shows that minimum--bias, DY and \jpsi distributions give redundant information as they can be related
in a simple way.

The main conclusion we can extract from the application of the model to the NA50 Pb--Pb data is that up to now there is no evidence
for \jpsi suppression strictly due to QGP formation and that the data can be well described by an absorption mechanism. However,
because of the density/opacity changes discussed above, it is not excluded that droplets of plasma could have been produced in the
Pb--Pb interaction. The only statement that can be made with certainty at present is that the drop in the ratio $J/\psi$ over DY is
due to absorption in a medium with increasing density and consequently with increasing opacity.

This naturally raises the question of whether one will ever be able to disentangle between \jpsi absorption and \jpsi destruction in a
deconfined medium. Only future experiments, in particular ALICE at the LHC, can enable us to answer this question. It is however
important to bear it in mind that any conclusions about deconfinement transitions using the mean path length $L$ in a nuclear medium can
lead to an incorrect interpretation of the behaviour of the observed results.

\bigskip\noindent
\textbf{\Large Acknowledgements}
\bigskip

Both authors express their thanks to Paula Bordalo and S\'ergio Ramos for providing access to the NA50 Pb--Pb
data and for very helpful discussions and clarifications concerning these data. This work was supported in part by contract 
PRAXIS/PCEX/P/FIS/124/96.

\newpage
\textbf{\Large Figure Captions}
\bigskip

\begin{description}
\item{\textbf{Fig.~1 --}} Experimental 1995 NA50 Pb--Pb data for the $E_T$ distribution of Drell--Yan $\mu^+\mu^--$ pairs \cite{tese}.
The solid curve corresponds to the fit by relation \eq{relation}.
\item{\textbf{Fig.~2 --}} Experimental 1995 NA50 Pb--Pb data for the $E_T$ distribution associated with the production of \jpsi \cite{tese}.
The solid curve corresponds to the fit by \eq{absorption} with $\epsilon=.6$.
\item{\textbf{Fig.~3.a --}} Experimental 1995 NA50 Pb--Pb data for the ratio \jpsi over Drell--Yan\cite{tese}. The solid curve is the fit obtained
by combining \eq{relation} with \eq{absorption} using $\epsilon=.6$.
\item{\textbf{Fig.~3.b --}} Combined 1995 and 1996 NA50 Pb--Pb data for the ratio \jpsi over Drell--Yan\cite{Ramello}. The solid curve is the fit obtained
by combining \eq{relation} with \eq{absorption} using $\epsilon=.6$.
\end{description}
\newpage
\begin{figure}
\begin{center}
\includegraphics{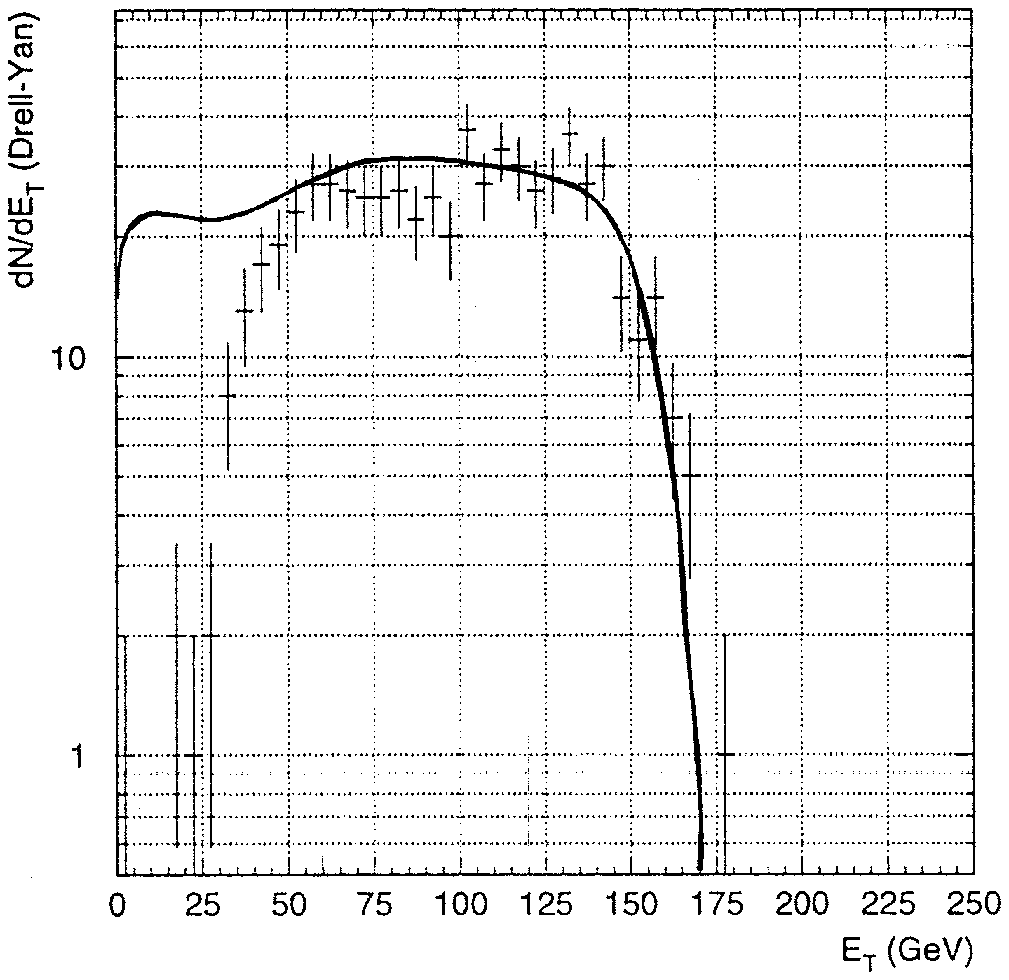}

\textbf{Fig. 1}
\end{center}
\end{figure}
\begin{figure}
\begin{center}
\includegraphics{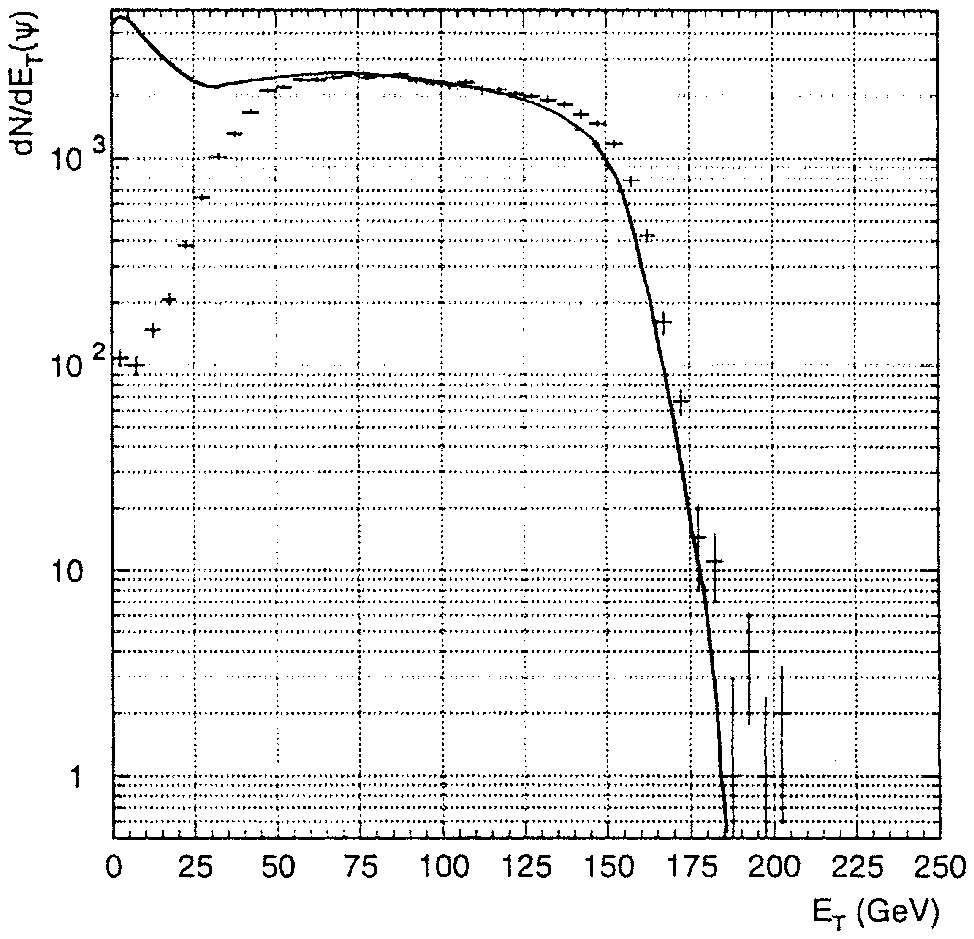}

\textbf{Fig. 2}
\end{center}
\end{figure}
\begin{figure}
\begin{center}
\includegraphics{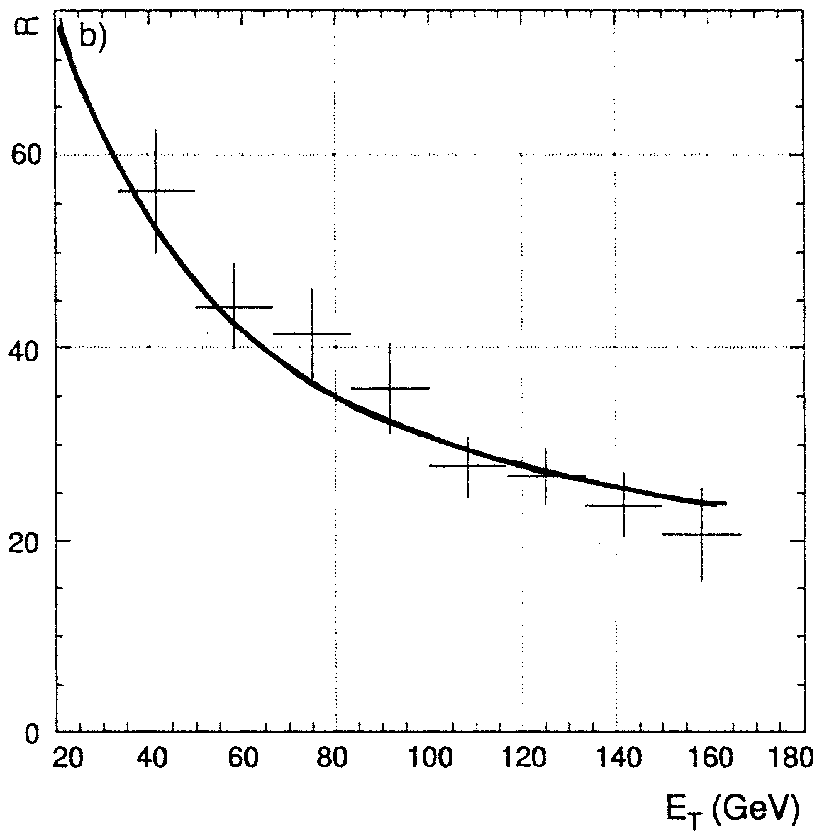}

\textbf{Fig. 3a}
\end{center}
\end{figure}
\begin{figure}
\begin{center}
\includegraphics{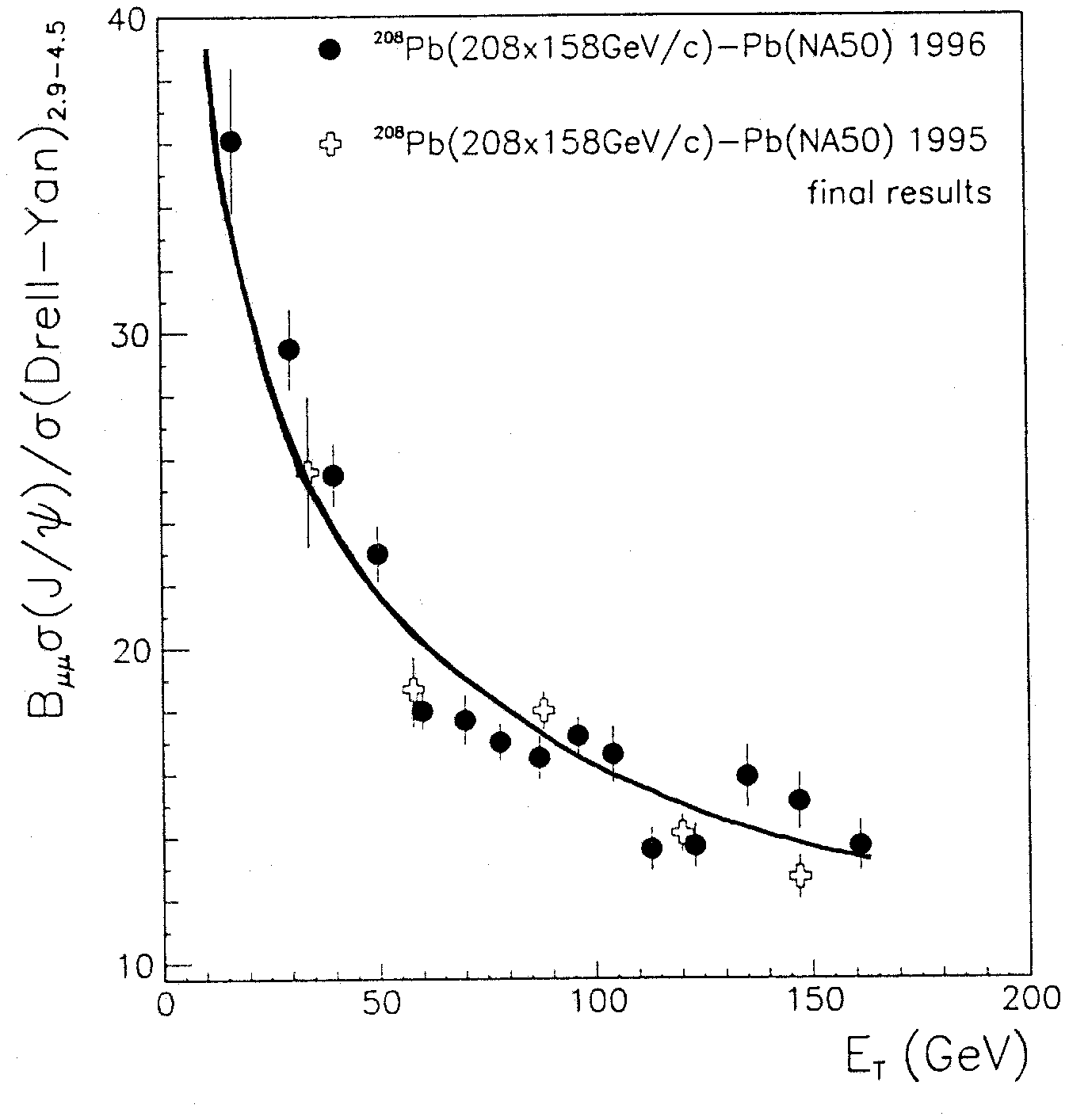}

\textbf{Fig. 3b}
\end{center}
\end{figure}
\end{document}